\documentclass[aip,rsi,reprint,graphicx,amssymb,amsmath]{revtex4-1} 

\usepackage{amssymb,amsmath}
\usepackage{graphicx}
\usepackage{bm}

\draft 

\newcommand{\muo}{ \mu_{\scriptscriptstyle 0}}

\begin{document}


\title{Passive Magnetic Shielding in Gradient Fields} 



\author{C.P. Bidinosti}
\email[Author to whom correspondence should be addressed. Electronic mail:]{c.bidinosti@uwinnipeg.ca}
\author{J.W. Martin}
\affiliation{Physics Department, The University of Winnipeg, 515
  Portage Avenue, Winnipeg, MB, R3B 2E9, Canada.}


\date{\today}

\begin{abstract}
The effect of passive magnetic shielding on dc magnetic field
gradients imposed by both external and internal sources is studied.
It is found that for concentric cylindrical or spherical shells of
high permeability material, higher order multipoles in the magnetic
field are shielded progressively better, by a factor related to the
order of the multipole.  In regard to the design of internal coil
systems for the generation of uniform internal fields, we show how one
can take advantage of the coupling of the coils to the innermost
magnetic shield to further optimize the uniformity of the field.
These results demonstrate quantitatively a phenomenon that was
previously well-known qualitatively: that the resultant magnetic field
within a passively magnetically shielded region can be much more
uniform than the applied magnetic field itself.  Furthermore we
provide formulae relevant to active magnetic compensation systems
which attempt to stabilize the interior fields by sensing and
cancelling the exterior fields close to the outermost magnetic
shielding layer.  Overall this work provides a comprehensive framework
needed to analyze and optimize dc magnetic shields, serving as a
theoretical and conceptual design guide as well as a starting point
and benchmark for finite-element analysis.

\end{abstract}


\maketitle 


\section{Introduction}

Passive magnetic shielding systems typically use a concentric
arrangement of thin shells of a high permeability material to divert
magnetic field lines around a region of interest.  The region within
the shielding system consequently possesses a reduced local magnetic
field.

While magnetic shielding is useful for a variety of applications, the
most stringent requirements are found in high precision experiments
where the limits of magnetometry technology are experienced or are
themselves being studied.  Some recent examples are in
biomagnetism~\cite{ptb,cohen}, electric dipole moment
experiments~\cite{brys,swallows}, and in developments of the most
precise atomic magnetometers~\cite{groeger,budkerromalis,romalis}.

Neutron electic dipole moment (EDM) experiments in particular suffer
from a systematic effect relating to the accrual of geometric phase as
neutrons and comagnetometer atoms sample the experimental volume
\cite{bib:gpe1,bib:gpe2,bib:gpe4,bib:gpe3}.  The geometric phase
effect is expected to present a dominant systematic effect in future
neutron EDM experiments.  To first approximation, the systematic
correction is proportional to the first-order gradient along the
direction of the applied magnetic field $\partial B_z/\partial z$.  It
is therefore important in these experiments both to limit and to
characterize magnetic field gradients.

While the analysis and development of single- and multi-layer magnetic
shields has been an important and active area of research for well
over a century~\cite{wills}$^{-}$\cite{paperno}, the focus in
analytical treatments has been almost exclusively on shielding uniform
magnetic fields.  To the best of our knowledge, only Urankar and
Oppelt~\cite{urankar} have explored the issue of passive magnetic
shielding in gradient fields from an analytical perspective.

Sumner {\it et al.}~\cite{sumner} provide an excellent overview of the
history of the field of magnetic shielding.  Exact solutions for
concentric cylindrical and spherical shields
\cite{wills,sterne,schweizer} have been simplified to approximate
formulae valid in the limit of high magnetic permeability and thin
shells~\cite{wadey,thomas,mager,freake,gubser,dubbers,sumner} as well
as to provide axial shielding factors for cylindrical geometries.
More recently, axial shielding in relation to shield spacing, end cap
holes, and gaps between mating surfaces has been explored
numerically~\cite{burt,paperno}.  Analytic treatments of the
quasi-static solutions have led to developments in external active
compensation~\cite{hoburg}.

However, as mentioned above, these authors considered only uniform
applied fields.  Urankar and Oppelt~\cite{urankar} analyzed the
general multipole field (both as an external and internal source) for
single spherical shields, and provided general shielding and reaction
factors. They employed their the results to analyze active magnetic
compensation used in conjunction with magnetically shielded rooms.
Quasi-static solutions valid in the dc limit were provided.  We extend
this work (in the dc limit) to multi-layer shields with spherical as
well as infinite cylindrical geometry.  For each, we consider the
following situations:
\begin{enumerate}
\item {\it Externally applied fields.}  Calculations are included both
  single and multi-layer shields.  The shielding factor for general
  multipole fields is calculated internal to the innermost shield and
  is of principal interest.  The field external to the outermost
  shield is also calculated, and is useful for designing active
  magnetic compensation systems. This field is dominated by the
  response of the outermost layer and the analysis is restricted to a
  single shield only.
\item {\it Internally applied fields.}  In many cases, such as in EDM
  experiments, a highly homogeneous internal field is desired and this
  is generally supplied by a coil system internal to a set of magnetic
  shields.  We consider here the impact of the innermost magnetic
  shield on general internally produced multipole fields, and
  calculate reaction factors by which the field internal to the coil
  system is amplified.
\end{enumerate}

We comment here on our primary new results:
\begin{itemize}
\item We report shielding factors, interior reaction factors, and
  exterior response fields for single layer, infinite cylindrical
  magnetic shields, exposed to general multipole dc applied fields.
  This extends the work of Ref.~\cite{urankar} from spherical to
  cylindrical geometries; for the spherical case, we demonstrate
  agreement with Ref.~\cite{urankar}.  Our results for single-layer
  shields are useful for designing active magnetic compensation
  systems (in the case of exterior response fields) and internal coil
  systems (in the case of interior reaction factors), and we provide
  useful examples of this.
\item We provide shielding factors for multi-layer shields in both
  cylindrical and spherical geometries for general multipole fields.
  One of our primary results is that higher multipole fields are
  always shielded better than the homogeneous field, a general result
  that should prove useful in applications requiring homogeneous
  fields.  This extends previous work to general multipole fields, and
  extends the work of Ref.~\cite{urankar} to multi-layer shielding
  systems in the dc limit.
\item Finally, we use a somewhat unique method of solution compared to
  previous authors, in that we consider the equivalent problem of
  bound surface currents.  While the end results are of course
  equivalent, our approach may be useful in certain situations.  We
  found, for example, that the consideration of surface currents gives
  a more direct conceptual link to the coils and current structures
  that one ultimately uses.
\end{itemize}

Our work is valid for dc fields, general multipole sources (both
internal and external to the magnetic shield), and any number of
concentric shields (cylindrical or spherical).  We provide an exact
treatment valid for shields of any permeability $\mu$ and thickness.
We also provide new approximate formulae in the high-$\mu$, thin shell
limit, which we have now validated for all higher multipoles.

We proceed first by describing our general method and then the
cylindrical and spherical applications of the method.  We conclude
with applications to some geometries of interest in EDM experiments,
which as noted above have very stringent requirements for magnetic
field quality.

\section{Problem statement and method of solution using equivalent bound surface currents}

Two problems of particular geometry are solved here using standard
cylindrical and spherical coordinates:
(i) the interaction of the transverse, 2-dimensional magnetic field
$\bm B=B_\rho(\rho,\phi) \bm{\hat{\rho}}+B_\phi(\rho,\phi)
\bm{\hat{\phi}}$ with infinitely-long cylindrical shells, and
(ii) the interaction of the general magnetic field $\bm
B=B_\rho(\rho,\theta,\phi) \bm{\hat{\rho}} +
B_\theta(\rho,\theta,\phi) \bm{\hat{\theta}} +
B_\phi(\rho,\theta,\phi) \bm{\hat{\phi}}$ with spherical shells.

As is commonly done to achieve analytic solutions for passive
shielding problems, we restrict our analysis to shields of linear,
homogeneous media, carrying no free current.  Under such conditions,
the response of a permeable object to an applied magnetic field can be
re-cast solely in terms of bound surface current $\bm K_b$ on the
surfaces of the object.  As a result, we take advantage of known
formulae for the magnetic fields of cylindrical and spherical sheet
currents~\cite{smythe,ferraro,lobb} to solve the appropriate boundary
conditions for sets of concentric magnetic shields.

For a shielding system of $M$ concentric shells, there are $2M$
distinct surface currents contributing to the net magnetic field in
each region.  Satisfying the boundary condition
 \begin{eqnarray}
   H^\parallel_{\rm in} &=&H^\parallel_{\rm out}\nonumber \\
&{\rm or}& \nonumber   \\
  \frac{1}{ \mu_{\rm in}} B^\parallel_{\rm in} &=&    \frac{1}{ \mu_{\rm out}} B^\parallel_{\rm out} 
  \label{bc}
\end{eqnarray}
for the tangential component of the magnetic field results in a set of
$2M$ simultaneous equations that determine the magnitudes of the
unknown surface currents.  By contrast, the typical means of solution
using the magnetic scalar potential (see e.g. Ref.~\cite{jackson})
gives a set of $4M$ simultaneous equations, albeit resulting in a
sparser matrix.

\section{The Infinitely Long Cylindrical Shield}

\subsection{The 2D multipole field generated by  a cylindrical current sheet}

From Refs.~\cite{smythe,ferraro,lobb}, an axial surface current
 \begin{equation}
\bm K = K \sin(n\phi) \, \bm{\hat{z}} 
\label{genK}
\end{equation}
with n-fold rotational symmetry ($n\geq 1$) bound to a cylindrical
surface $\rho =a$ gives rise to the vector potential
\begin{equation}
\bm A = \mathcal{K}\, \frac{\sin(n\phi)}{n}
\begin{cases}
\rho^{n}  \, \bm{\hat{z}} &  \rho < a   \\[.1cm]
\dfrac{a^{2n}}{\rho^{n}} \, \bm{\hat{z}}  &  \rho > a 
\end{cases}
\, ,
\label{Acyl}
\end{equation}
where $\mathcal{K} =\muo K/(2a^{\,n-1})$ has units T/m$^{n-1}$.  The
introduction of $\mathcal{K}$, while not necessary, leads to a
simplified notation for the determination of shielding factors,
especially when multiple shields are considered.  The magnetic field
arising from Eq.~\ref{Acyl} is
\begin{equation}
\bm B = \mathcal{K}
\begin{cases}
\rho^{n-1} \, [\cos(n\phi) \, \bm{\hat{\rho}} - \sin(n\phi) \, \bm{\hat{\phi}}] &  \rho < a   \\[.1cm]
\dfrac{a^{2n}}{\rho^{n+1}} \, [\cos(n\phi) \, \bm{\hat{\rho}} + \sin(n\phi) \, \bm{\hat{\phi}} ]  &  \rho > a 
\end{cases}
\, .
\label{Bcyl}
\end{equation}
We use these results to solve the following problems.

\subsection{A single cylindrical shield in an external field}

Consider an infinitely-long cylindrical shield of inner radius $R$,
thickness $t$, and permeability $\mu$ in the presence of an externally
applied transverse magnetic field
\begin{equation}
 \bm B_{\rm ext} = G_n \, \rho^{n-1} \, [\cos(n\phi) \, \bm{\hat{\rho}} - \sin(n\phi) \, \bm{\hat{\phi}} ] 
 \label{Bext}
 \end{equation}
with a magnitude gradient $G_n$ in T/m$^{n-1}$.  The case $n=1$
therefore corresponds to a uniform field, and $n>1$ corresponds to
higher multipoles.  By symmetry, the bound currents induced on the
inner surface ($r_1=R$) and outer surface ($r_2=R+t$) of the magnetic
shield have the same harmonic $n$ as $B_{\rm ext}$ and generate fields
given by Eq.~\ref{Bcyl}.

To find the coefficients $\mathcal{K}_1$ and $\mathcal{K}_2$,
representative of the bound surface current on the inner and outer
surfaces of the shield, respectively, the boundary condition of
Eq.~\ref{bc} is applied to the azimuthal components $B_\phi$ of the
magnetic field.  This results in the following system of equations:
\begin{equation}
(\mu+\muo)\,  \mathcal{K}_1 + (\mu-\muo)\,\mathcal{K}_2 = - (\mu-\muo) G_n
\label{Keq1}
\end{equation}
\begin{equation}
(\mu-\muo) \left( \frac{r_1}{r_2} \right)^{2n} \mathcal{K}_1 + (\mu+\muo) \, \mathcal{K}_2 = (\mu-\muo)G_n \, ,
\label{Keq2}
\end{equation}
which have solutions
\begin{equation}
\mathcal{K}_1= -2G_n \, \frac{ \mu (\mu-\muo)}{(\mu+\muo)^2-
  (r_1/r_2)^{2n} \, (\mu-\muo)^2}
 \label{K1}
\end{equation}
and 
\begin{equation}
 \mathcal{K}_2= G_n \, \frac{\mu^2-\muo^2+ (r_1/r_2)^{2n} \, (\mu-\muo)^2}{(\mu+\muo)^2- (r_1/r_2)^{2n} \, (\mu-\muo)^2} \, \cdot
  \label{K2}
\end{equation}

Defining the shielding factor $S$ as the applied field divided by the
net internal field~\cite{thomas, mager, dubbers, sumner} gives
\begin{eqnarray}
S &=& \frac{G_n}{ \mathcal{K}_1+ \mathcal{K}_2+ G_n}  \nonumber \\
 &&\nonumber \\
&=& \frac{ (\mu+\muo)^2-(r_1/r_2)^{2n} \, (\mu-\muo)^2}{4\mu \muo} \\
 && \nonumber \\
&=&  1+ \frac{ (\mu-\muo)^2}{4\mu \muo} \left[ 1-\left(\frac{r_1}{r_2}\right)^{2n}\right]\ \, .
\label{Scyl}
\end{eqnarray}
In the limit $R \gg t$ and $\mu \gg \muo$, this reduces to
\begin{equation}
S \simeq 1+ \frac{ \mu}{\muo} \frac{n}{2}\frac{t}{\bar{R}}\, ,
\label{Scylthin}
\end{equation}
where $\bar{R}=R+t/2$ is the average radius of the shield.  

The results of Eqs.\ref{Scyl} and \ref{Scylthin} for the $n=1$ case
(i.e., a uniform applied field) agree with previous
work.\cite{kaden,nussbaum, durand, wills,thomas, mager, dubbers,
  sumner, hoburg} The important new result here is the generalization
to higher $n$, where we find that higher multipole fields are always
shielded better than $n=1$ case.  In the thin shield limit, in
particular, the shielding factor increases proportional to $n$.
Taking a linear combination of external fields, summing Eq.~\ref{Bext}
over $n$, we would therefore find that the interior shielded volume
always becomes more uniform, i.e. the higher multipoles are suppressed
more strongly.

We now consider the {\it exterior} field $\bm B_{\rm shield}$, defined
as the additional field induced by the presence of the magnetic
shield.  An important consideration for active shielding systems
(which feed back on measurements of the net magnetic field outside the
passive shield assembly) is the perturbation $\bm B_{\rm shield}$
superimposed on the applied field in the region $\rho>r_2$.  From
Eqs.~\ref{Bcyl}, \ref{K1} and \ref{K2}, the general solution is
\begin{equation}
\bm B_{\rm shield}= \frac{\mathcal{K}_1\, r_1^{2n}+ \mathcal{K}_2 \, r_2^{2n} }{\rho^{n+1}} \, [\cos(n\phi) \, \bm{\hat{\rho}} + \sin(n\phi) \, \bm{\hat{\phi}} ] \, ,
\end{equation}
which  for $\mu \gg \muo$ reduces to
\begin{eqnarray}
\bm B_{\rm shield}=  G_n \frac{r_2^{2n}}{\rho^{n+1}} \, [\cos(n\phi) \, \bm{\hat{\rho}} + \sin(n\phi) \, \bm{\hat{\phi}} ] \, .
\end{eqnarray}
This in turn can be recast as
\begin{eqnarray}
\bm B_{\rm shield}= \frac{\muo}{4\pi}  \frac{m'_n}{\rho^{n+1}} \, [\cos(n\phi) \, \bm{\hat{\rho}} + \sin(n\phi) \, \bm{\hat{\phi}} ] \, ,
\end{eqnarray}
where $m'_n=4\pi G_nr_2^{2n}/\muo $ is the $(n+1)^{\rm th}$ multipole
moment per unit length defined by
\begin{equation}
\bm A = 
\dfrac{\muo\, m'_n}{4\pi} \dfrac{\sin(n\phi)}{n \, \rho^n} \, \bm{\hat{z}}
\end{equation}
for the vector potential outside a current-carrying cylinder from
Eq.~\ref{Acyl}.

The result for the exterior field is important because it may also be
applied to multi-layer shielding systems, since it is the response
of the outermost shield that dominates.  Furthermore, as in the
magnetic shielding case above, the exterior field may be decomposed
into multipoles.  The results can therefore be used to decide the
optimal placement of the magnetic sensors in the active magnetic
compensation system, because the sensors can be placed selectively to
accentuate sensitivity to particular multipoles, considering also the
steeper suppression of higher multipoles with increasing $\rho$.

\subsection{Multiple shields in an external field}

\begin{figure}[hbt]
\centering
\includegraphics[keepaspectratio, width=3.in]{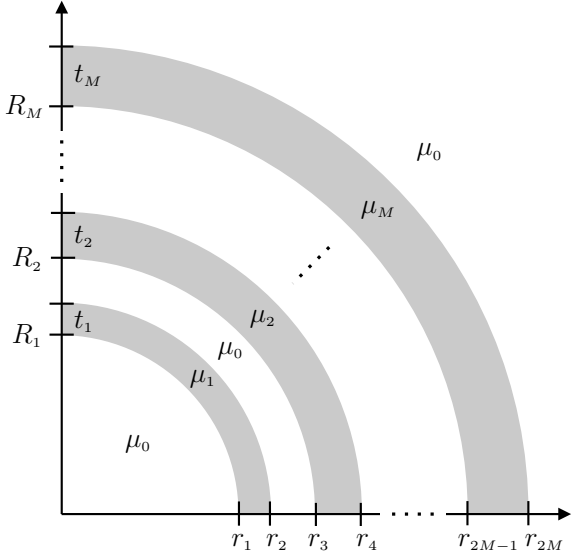}
\caption{Cross-sectional view (first quadrant) of $M$ concentric
  cylindrical or spherical shells separated by free space.  The
  material boundaries are located at radial positions
  $r_{\scriptscriptstyle 1}$ through $r_{\scriptscriptstyle 2M}$.  The
  inner radius $R$, thickness $t$, and permeability $\mu$ of each
  shield is indicated on the drawing. }
\label{Shields}
\end{figure}

Now consider a set of $M$ concentric cylinders in an applied external
field given by Eq.~\ref{Bext}.  The geometry is shown in
Fig.~\ref{Shields}, where our conventions for labelling are also
described.  The $m$-th cylinder has an inner radius $R_m$, an outer
radius $R_m+t_m$, a thickness $t_m$ and a permeability $\mu_m$.  There
are now $2M$ bound surface currents that one must find.  The $i$-th
surface current $\mathcal{K}_i$ resides on the {\it inner} surface of
the $m$-th shield if $i$ is odd (i.e., $i=2m-1$) and on its {\it
  outer} surface if $i$ is even (i.e. $i=2m$).  The radial position of
$\mathcal{K}_i$ is thus defined as
\begin{equation}
r_i=
\begin{cases} 
R_m & \mbox{for $i=$ odd  and $m=\tfrac{i+1}{2}$ }  \\
R_m + t_m &\mbox{for $i=$ even  and $m=\tfrac{i}{2}$,} 
\end{cases} 
\label{radii}
\end{equation}
i.e. $r_1=R_1$ is the inner surface of the innermost shield,
$r_2=R_1+t_1$ is the outer surface of the innermost shield, $r_3=R_2$
is the inner surface of the next-to-innermost shield, and so on.

Satisfying the boundary condition of Eq.~\ref{bc} at each surface
leads to the general system of equations
\begin{equation}
\bm A \bm{\mathcal{K}} = G_n \bm I \, ,
\label{AKG}
\end{equation}
where   $\bm I=[1,1, \dots, 1]^{\rm T}$, $\bm{\mathcal{K}}=[\mathcal{K}_1, \mathcal{K}_2, \dots, \mathcal{K}_{2M}]^{\rm T}$, and $\bm A$ is a $2M \times 2M$ matrix with elements
\begin{equation}
a_{ij}= 
\begin{cases} 
(r_j/r_i)^{2n}&\mbox{for } j<i \\
U_{m} &\mbox{for  $j=i=$ odd and $m=\tfrac{i+1}{2}$}  \\ 
 V_{m} &\mbox{for  $j=i=$ even and  $m=\tfrac{i}{2}$ }  \\ 
-1&\mbox{for } j>i 
\label{Aelements}
\end{cases} 
\, ,
\end{equation}
where
\begin{equation}
U_m = -V_m=-\frac{\mu_m+\muo}{\mu_m-\muo} \, ,
\end{equation}
and $r_i$ and $r_j$ are defined per Eq.~\ref{radii}.  The $2\times2$
diagonal submatrices of $\bm A$ correspond to Eqs.~\ref{Keq1}
and~\ref{Keq2} for each individual, isolated shield.  To illustrate,
the explicit form of the general matrix $\bm A$ for $M=2$ shields is
\begin{equation}
\bm A =
 \begin{pmatrix}
-\frac{\strut \mu_1+\muo}{\strut \mu_1-\muo}	& -1 		& -1	 	& -1	  	 \\[12pt]
\left(\! \frac{\strut R_1}{\strut R_1+t_1} \! \right)^{\! \! 2n} 	&\frac{\strut \mu_1+\muo}{\strut  \mu_1-\muo}	& -1  		&-1	\\[12pt]
\left(\!  \frac{\strut  R_1}{R_2}\!  \right)^{\! \! 2n} & \left(\!  \frac{\strut  R_1+t_1}{\strut  R_2}\!  \right)^{\! \! 2n}	&	- \frac{\strut  \mu_2+\muo}{\strut \mu_2-\muo} & -1 	 \\[12pt]
\left(\!  \frac{\strut  R_1}{\strut R_2+t_2}\!  \right)^{\! \! 2n}	 & \left(\!  \frac{\strut R_1+t_1}{\strut R_2+t_2}\! \right)^{\! \! 2n}	& \left(\!  \frac{\strut  R_2}{\strut R_2+t_2}\! \right)^{\! \! 2n} & \frac{\strut  \mu_2+\muo}{\strut \mu_2-\muo} 
 \end{pmatrix} \, . \nonumber
 \label{Amatrix2}
\end{equation}

Returning now to the general case, the total combined shielding factor
for $M$ concentric cylindrical shields is
\begin{equation}
S_{\rm tot} =G_n   \left(G_n+ \sum_{i=1}^{2M}\mathcal{K}_i \right)^{-1} \, ,
\label{generalS}
\end{equation}
with the $\mathcal{K}_i$ determined from Eqs.~\ref{AKG}
and~\ref{Aelements}.  An algebraic scheme for the solution of
Eq.~\ref{generalS} is given by Wills~\cite{wills}, and one can show
that our result agrees with his explicit formulation of $S_{\rm tot}$
for double and triple cylindrical shields of the same permeability $\mu$.  With the generating
formulae of Eq.~\ref{Aelements}, Eq.~\ref{AKG} is also readily coded
and solved using any number of computer programs designed for symbolic
or numeric computation.~\cite{code}

The results of solving these equations for multi-layer shields, along
with approximate formulae valid for the small-$t$, high-$\mu$ limit,
will be discussed in Section~\ref{sec:results}.  A key generic feature
will be the suppression of higher $n$ in powers of the number of
shields.

\subsection{A single cylindrical shield with an internal coil}
\label{SecD}

We now turn to the study of coil systems internal to the magnetic
shield system.  In this case, the modification of the internal field
is dominated by the innermost magnetic shield.  We therefore consider
a single cylindrical shield in order to simplify the discussion.

Consider an applied surface current $\bm K$ of the form Eq.~\ref{genK}
on a coaxial cylindrical surface $\rho=a$ inside a single shield of
inner radius $r_1=R$, outer radius $r_2=R+t$, and permeability $\mu$.
Solving boundary conditions gives the following system of equations:
\begin{equation}
(\mu-\muo)\left( \frac{a}{r_1} \right)^{2n}  \mathcal{K}_a - (\mu+\muo)\,  \mathcal{K}_1 - (\mu-\muo)\,\mathcal{K}_2 = 0
\end{equation}
\begin{equation}
(\mu-\muo)\left( \frac{a}{r_2} \right)^{2n}  \mathcal{K}_a+(\mu-\muo)\left( \frac{r_1}{r_2} \right)^{2n}  \mathcal{K}_1 + (\mu+\muo)\,\mathcal{K}_2 = 0 \, ,
\end{equation}
where $\mathcal{K}_a=\muo K/(2a^{\,n-1})$.  The equations are again solved for $\mathcal{K}_1$ and $\mathcal{K}_2$.

The ratio of field in the region $\rho<a$, divided by the field
without the shield may then be calculated.  We call this ratio the
reaction factor $C$, in keeping with the terminology of
Ref.~\cite{urankar}.  The result for the reaction factor is
\begin{eqnarray}
C & =&  \frac{\mathcal{K}_a+\mathcal{K}_1+\mathcal{K}_2}{\mathcal{K}_a} \\
\label{Cgen}
&=&1 +  \left(\frac{a}{r_1}\right)^{\!2n}\,\frac{(\mu - \muo)\, (\mu +\muo) \, \gamma_n }{4\mu \muo +(\mu - \muo)^2 \,\gamma_n}
\end{eqnarray}
where $\gamma_n\equiv 1-(r_1/r_2)^{2n}$.  In the limit $\mu \gg \muo$
this reduces to
\begin{equation}
C = 1+ \left( \frac{a}{R} \right)^{2n} \, ,
\label{Ccyl}
\end{equation}
and one sees that the internal field is augmented more strongly for
small $n$ than it is for large $n$ because $a<R$.  In the limit $a=R$,
the reaction factor is identically 2, independent of $n$.

These results are applied to a sample internal coil design in
Sec.~\ref{sec:results}.  A key feature will again be that fields are
in general more homogeneous with the shield than without, but that
optimal homogeneity can be achieved for a particular geometrical
factor $a/R$.

\section{The spherical shield}
\label{SecSphere}

\subsection{The zonal multipole field generated by a spherical  current sheet}

In general, a surface current bound to a sphere, and its resulting
magnetic field, can be written in terms of spherical harmonics of
order $m$ and degree $n$~\cite{smythe,ferraro}.  One can show,
however, that the resulting equations arising from the boundary
conditions on the tangential components of the magnetic field (i.e.,
$B_\theta$ and $B_\phi$) are independent of the order $m$ of the
spherical harmonic.  Without loss of generality, we can therefore
restrict the analysis of spherical shields to zonal surface currents
and fields only (i.e., $\phi$-independent, $m=0$), a simplification
also noted by Urankar and Oppelt~\cite{urankar}.

The results for a given $n$ can therefore be applied, without loss of
generality, to cases where tesseral components (i.e., $m>0$) exist in
the fields and currents.  This is extremely valuable from the point of
view of coil design, where the general spherical harmonics can used as
\textit{building blocks} to produce a desired magnetic field~\cite{romeo}.

From Refs.~\cite{smythe,ferraro}, the zonal surface current
\begin{equation}
\bm K = K \, P_n^1(u) \, \bm{\hat{\phi}}
\label{Ksphere}
\end{equation}
bound to a spherical surface $r =a$ gives rise to the vector potential
\begin{equation}
\bm A = \mathcal{K}\,P_n^1(u)
\begin{cases}
r^{n}  \, \bm{\hat{\phi}} &  r < a    \\[.1cm]
\dfrac{a^{2n+1}}{r^{n+1}} \, \bm{\hat{\phi}}  &  r > a 
\end{cases}
\, ,
\label{Asphere}
\end{equation}
where $P_n^1(u) $ is the associated Legendre function of order 1 and
degree $n$, $u=\cos\theta$ , and the coefficient $\mathcal{K} =\muo
K/((2n+1)a^{n-1})$ has units T/m$^{n-1}$.  The magnetic field arising
from Eq.~\ref{Asphere} is
 \begin{equation}
\bm B = \mathcal{K}
\begin{cases}
 r^{n-1} \, (n+1)\,[n P_n(u) \, \bm{\hat{r}} -  P_n^1(u)  \, \bm{\hat{\theta}} ] &  r < a   \\[.1cm]
\dfrac{a^{2n+1}}{r^{n+2}} \, n \,[(n+1) P_n(u) \, \bm{\hat{r}} +  P_n^1(u)  \, \bm{\hat{\theta}} ]  &  r > a 
\end{cases}
\, ,
\end{equation}
where $P_n(u)$ is the Legendre function of degree $n$.
We use these results to solve the following problems.

\subsection{A single spherical shield in an external field\label{sec:singlesphere}}

Consider a spherical shield of inner radius $r_1=R$, outer radius
$r_2=R+t$, and permeability $\mu$ in the presence of an externally
applied magnetic field
\begin{equation}
\bm B_{\rm ext} = G_n \, r^{n-1} \, [n(n+1) P_n(u) \, \bm{\hat{r}} -  (n+1) P_n^1(u)  \, \bm{\hat{\theta}} ] 
\label{BextS}
\end{equation}
with a magnitude gradient $G_n$ in  T/m$^{n-1}$.  The
method of analysis follows exactly as above, and the solution of the
boundary conditions on the tangential field $B_\theta$ leads to the
general shielding factor
\begin{equation}
S=1+ \frac{ (\mu-\muo)^2}{\mu \muo} \frac{n(n+1)}{(2n+1)^2} \left[ 1-\left(\frac{r_1}{r_2}\right)^{2n+1}\right]\, ,
\label{Ssphere}
\end{equation}
which agrees with Ref.~\cite{urankar}.
In the limit  of
  a thin shield ($t<<R$) with large permeability $(\mu>>\muo)$,  
  the shielding factor can be approximated as
\begin{equation}
S \simeq 1+ \frac{ \mu}{\muo}\frac{n(n+1)}{2n+1} \frac{t}{\bar{R}}\, .
\label{Sspherelthin}
\end{equation}
The results of Eqs.~\ref{Ssphere} and~\ref{Sspherelthin} for the $n=1$
case (i.e., a uniform applied field) agree with previous
authors~\cite{jackson,wills,schweizer}.  Similar to the cylindrical
case, higher $n$ fields are shielded progressively better, and in the
large-$n$ limit the shielding factor again becomes proportional to
$n$.  In cases where the applied magnetic field would be a linear
combination of fields with differing $n$, the magnetic field internal
to the magnetic shield would therefore always be more uniform than the
applied field, i.e., higher multipoles are suppressed more strongly.

We now again consider the {\it exterior} field induced by the presence
of the magnetic shield in the region $r>r_2$.  In this case, the
perturbation of the external field by a spherical shield of $\mu \gg
\muo$ is
\begin{eqnarray}
\hspace{-.5cm}
\bm B_{\rm shield}
 &= & G_n (n+1) \frac{r_2^{2n+1}}{r^{n+2}}  \,  [(n+1) P_n(u) \, \bm{\hat{r}} +  P_n^1(u)  \, \bm{\hat{\theta}} ] \nonumber \\[.1cm]
 &=& \frac{\muo}{4\pi} \frac{m_n}{r^{n+2}} \, [n(n+1) P_n(u) \, \bm{\hat{r}} + n P_n^1(u)  \, \bm{\hat{\theta}} ]  \, ,
\end{eqnarray}
where $m_n=4\pi G_nr_2^{2n+1} (n+1)/(n\muo) $ is the $(n+1)^{\rm th}$
multipole moment defined by
\begin{equation}
\bm A = 
\dfrac{\muo\, m_n}{4\pi \, r^{n+1}}\, P_n^1(u)   \, \bm{\hat{\phi}}
\end{equation}
for the vector potential outside a current-carrying sphere from
Eq.~\ref{Asphere}.

\subsection{Multiple spherical shields in an external field\label{sec:multiplespheres}}

For $M$ concentric shields, we again have the same system of equations
$\bm A \bm{\mathcal{K}} = G_n \bm I$ where now the general matrix
elements of $\bm A$ are
\begin{equation}
a_{ij}= 
\begin{cases} 
\tfrac{n}{n+1}(r_j/r_i)^{2n+1}&\mbox{for } j<i \\
U_{m} &\mbox{for  $j=i=$ odd and $m=\tfrac{i+1}{2}$}  \\ 
 V_{m} &\mbox{for  $j=i=$ even and  $m=\tfrac{i}{2}$ }  \\ 
-1&\mbox{for } j>i 
\label{AelementsS}
\end{cases} 
\, ,
\end{equation}
with
\begin{equation}
U_m = - \frac{(n+1) \, \mu_m+n \, \muo}{(n+1)(\mu_m-\muo)} \, ,
\end{equation}
\begin{equation}
V_m = \frac{n\, \mu_m+(n+1)\, \muo}{(n+1)(\mu_m-\muo)} \, ,
\end{equation}
and $r_i$ and $r_j$ are defined per Eq.~\ref{radii}.  In general
the total combined shielding factor for $M$ concentric spherical
shields is given by Eq.~\ref{generalS} with the $\mathcal{K}_i$
determined from Eqs.~\ref{AKG} and~\ref{AelementsS}.

One can show that the general shielding factor of Eq.~\ref{generalS}
reduces to the explicit formula for double and triple spherical
shields of the same permeability.~\cite{wills, schweizer}
 
We calculate sample results for multi-layer magnetic shields in
Section~\ref{sec:results}.  Similar to the cylindrical case, a generic
feature will be a suppression of higher $n>1$ and a more uniform
resultant internal shielded field.

\subsection{A single spherical shield with an internal coil}

Again driven by the desire to create an internal coil system that is
homogeneous, we consider internal coils wound on a spherical surface
inside the magnetic shielding system.  As in the cylindrical case, the
modification of the internal field will be dominated by the response
of the innermost magnetic shield, and we restrict the analysis to a
single spherical shield.

Consider an applied surface current $\bm K$ of the form
Eq.~\ref{Ksphere} on $r=a$ inside a spherical shield of inner radius
$r_1=R$, outer radius $r_2=R+t$, and permeability $\mu$.  Following
the method laid out in Sec.~\ref{SecD}, the reaction factor giving the
ratio of field in the region $r<a$ with and without the shield is
\begin{eqnarray}
C &= &1 + \left(\frac{a}{r_1}\right)^{2n+1} \nonumber \\
  &&\times \, \dfrac{n(\mu - \muo)\, (n(\mu +\muo)+\muo) \, \gamma_n }{(2n+1)^2\mu \muo +n(n+1)(\mu - \muo)^2\, \gamma_n} \, ,
\end{eqnarray}
where now $\gamma_n=1-(r_1/r_2)^{2n+1}$.
In the limit $\mu \gg \muo$ this reduces to
\begin{equation}
C = 1+ \frac{n}{n+1} \, \left( \frac{a}{R} \right)^{2n+1} \, .
\label{Csphere}
\end{equation}
These results agree with Ref.~\cite{urankar}.  An interesting
difference with the cylindrical case is the prefactor $n/(n+1)$
preceding the second term.  Because of it, there is a cross-over
behaviour in the relative magnitudes of the reaction factors and one finds that
higher order fields become augmented more strongly (not less) by the presence of the
shield as  $a/R \rightarrow 1$.  This is discussed further in
Sec.~\ref{internalCoils}.

\section{Results and Applications \label{sec:results}}

\subsection{Multiple shields: Numerical results and useful approximations}
 
Most practical interests lie in the construction of multiple shields
made of thin material ($t_m \ll R_m$) with large permeability
($\mu_m\gg \muo$).  Many previous authors provided approximations for
designing shields in this regime.  A well-known result, for the total
shielding factor $S_{\rm tot}$ for well-separated
shields~\cite{thomas, freake, sumner}, is generalized to any $n$ as
follows:
\begin{equation}
S_{\rm tot} \simeq   \prod_{m=1}^{M-1} S_M \, S_m \left[ 1- \left( \frac{\bar{R}_m}{\bar{R}_{m+1}}\right)^{\!\beta}\, \right] \, ,
\label{far}
\end{equation}
where $S_m$ is the shielding factor of the $m$-th shield (from
Eq.~\ref{Scylthin} or~\ref{Sspherelthin}), $\bar{R}_m$ is the average
radius of the $m$-th shield, and the exponent $\beta$ equals $2n$ for
cylinders and $2n+1$ for spheres.

\begin{figure}[hbt]
\centering
\includegraphics[keepaspectratio, width=3.2in]{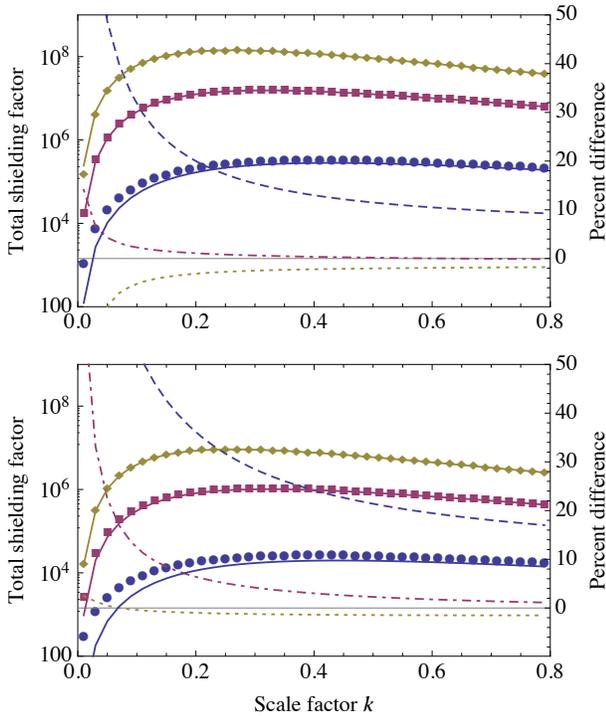}
\caption{The total shielding factor of four concentric cylindrical
  shells of permeability $\mu/\muo=4\times10^4$ (top) and
  $2\times10^4$ (bottom) determined from Eq.~\ref{generalS} for an
  applied field of $n=1$ (blue circles), 2 (red squares), and 3
  (yellow diamonds).  The solid lines are the results of
  Eq.~\ref{far}.  The right ordinate axis gives the percent difference
  between Eqs.~\ref{generalS} and \ref{far} for $n=1$ (dashed line), 2
  (dot-dashed line), and 3 (dotted line). }
\label{ResultsCyl}
\end{figure}

\begin{figure}[hbt]
\centering
\includegraphics[keepaspectratio, width=3.2in]{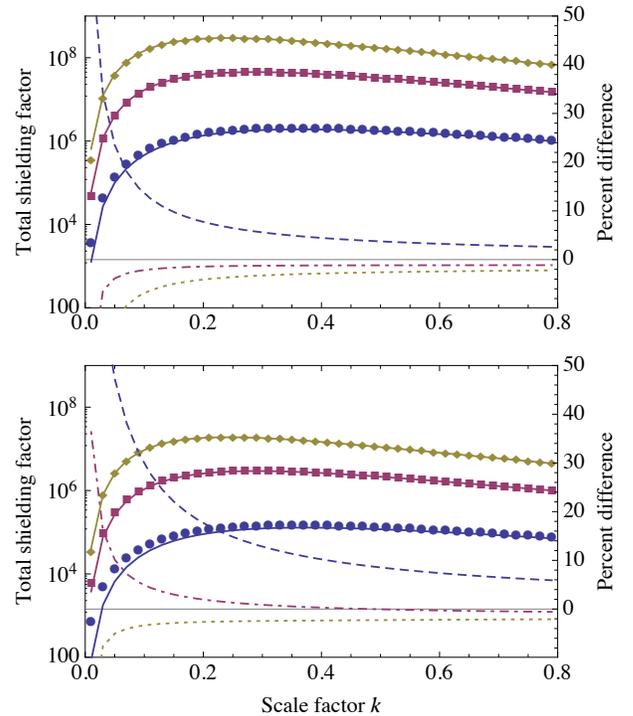}
\caption{The total shielding factor of four concentric spherical
  shells of permeability $\mu/\muo=4\times10^4$ (top) and
  $2\times10^4$ (bottom) determined from Eq.~\ref{generalS} for an
  applied field of $n=1$ (blue circles), 2 (red squares), and 3
  (yellow diamonds).  The solid lines are the results of
  Eq.~\ref{far}.  The right ordinate axis gives the percent difference
  between Eqs.~\ref{generalS} and \ref{far} for $n=1$ (dashed line), 2
  (dot-dashed line), and 3 (dotted line). }
\label{ResultsSphere}
\end{figure}

In Figs.~\ref{ResultsCyl} and~\ref{ResultsSphere} we compare
Eq.~\ref{far} with the general result of Eq.~\ref{generalS} for
cylindrical and spherical shields, respectively.  We analyze a shield
geometry that is likely typical of many applications: four concentric
shields each of the same thickness $t=\tfrac{1}{16}$~inches $\sim
1.6$~mm (a standard size) with a radius $R_1=0.5$~m for the inner most
shield.  All shields have the same permeability and we examine two
specific cases: $\mu=2\times10^4\, \muo$ and $\mu=4\times10^4\, \muo$.
The shield spacings are set by a single geometrical scale factor $k$,
such that the inner radius of the $m$-th shield is $R_m=(1+k)^{m-1} R_1$.

A key feature is that higher $n$ are always progressively suppressed
as $n$ increases.  For example, for the four-layer shield explored
here, the shielding factor for $n=2$ is of order $10^2$ greater than
for $n=1$.  The optimal choice of scale factor $k$ is relatively
independent $n$.

Furthermore, the approximate formulae of Eq.~\ref{far} appear to be
even more accurate for higher $n$ than for the $n=1$ case.  This is
shown in Figs.~\ref{ResultsCyl} and~\ref{ResultsSphere} as a percent
difference from the exact result.

\begin{figure}[hbt]
\centering
\includegraphics[keepaspectratio, width=3.2in]{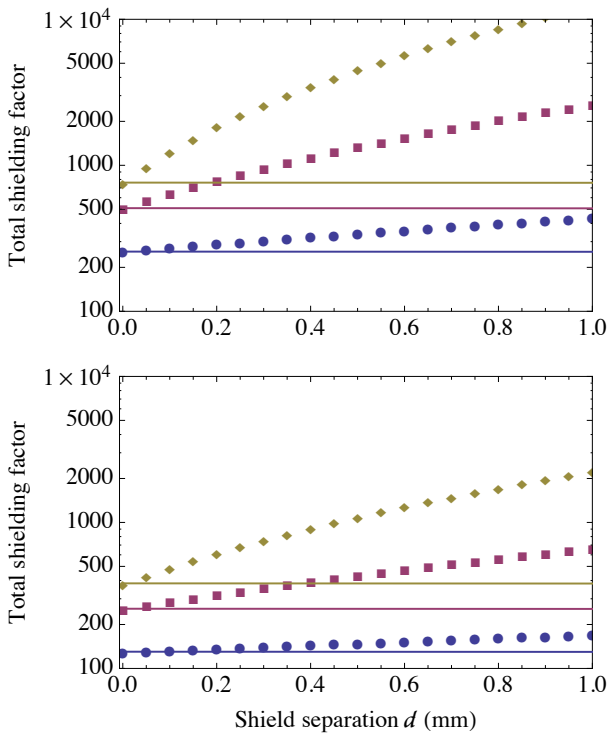}
\caption{The total shielding factor of four closely-spaced concentric
  cylindrical shells of permeability $\mu/\muo=4\times10^4$ (top) and
  $2\times10^4$ (bottom) determined from Eq.~\ref{generalS} for an
  applied field of $n=1$ (blue circles), 2 (red squares), and 3
  (yellow diamonds).  The solid lines are the results of
  Eqs.~\ref{close} and~\ref{Scylthin}, and are a very weak inverse
  function of $d$.}
\label{ResultsCylPacked}
\end{figure}

For closely packed  cylindrical and spherical shields,  on the other hand,
a useful approximation for the total
shielding factor is
\begin{equation}
S_{\rm tot} \simeq \sum_{m=1}^M S_m \, ,
\label{close}
\end{equation}
which is now validated for all $n$.  For shields that just touch,
Eq.\ref{close} correctly approximates the shielding factor of a single
shield with thickness equivalent to the total thickness of the
shielding material.  As an example, we show in
Fig.~\ref{ResultsCylPacked} plots of $S_{\rm tot}$ as a function of a
small separation $d$ between each of the four concentric cylindrical
shields discussed above.  Similar results hold for spherical shields.

At $d=0$, we find that for the range of parameters studied here
Eq.~\ref{close}, using Eq.~\ref{Scylthin} for the $S_m$, over predicts
$S_{\rm tot}$ by $\sim 3-5$\% compared to the exact result of
Eq.~\ref{generalS}.  This is reduced slightly if one uses
Eq.~\ref{Scyl} for the $S_m$.  We also point out, that as expected,
the value of $S_{\rm tot}$ from Eq.~\ref{generalS} for the four
shields with $d=0$ agrees exactly with Eq.~\ref{Scyl} for a single
shield that is four times as thick.

The results of Fig.~\ref{ResultsCylPacked} also highlight the
importance of sufficiently separating the shells.  An interesting
observation is that the shielding of higher order $n>1$ fields
increases dramatically with $n$ with even sub-millimetre shield
spacing.  This may ague for subdividing shields further, possibly with
thin interstitial nonmagnetic layers, if desiring particularly to
reduce gradients with relatively less impact on the uniform field
case.  For example, an application requiring better control of $n>1$
could use four well-separated shields to reduce $n=1$, but each of
those four shields could be thinner layers separated by a thin plastic
layer to relatively augment the shielding of $n>1$.

\subsection{The external physical dipole}

The source of external gradient fields can often be linked to some
nearby \textit{dipole} -- a research magnet, a steel door, or even a
passing vehicle~\cite{brys}.  A very important example to study then
is the field of the physical dipole, or current loop, expressed in
spherical coordinates.  More complicated magnetic structures can often
be modelled from a superposition of such loops or, as mentioned
before, using a decomposition into general spherical
harmonics.~\cite{romeo}
 
Here we consider a circular loop of radius $r_c$ carrying current $I$ that is
co-axial with $z$-axis and lying in the plane $z=z_c$. The loop can
also be viewed as lying on a sphere of radius $a=\sqrt{r_c^2+z_c^2}$
at the polar angle $\alpha=\tan^{-1}{r_c/z_c}$. The magnetic field of
the loop can be decomposed into zonal harmonics~\cite{smythe,ferraro}
and therefore its interaction with spherical shields is easily
determined using the results of Sec.~\ref{SecSphere}.  

For example, in the region $r<a$ the magnetic field components of the
loop are
\begin{eqnarray}
\left( 
\begin{array}{c}
B_r \\[.2cm]
B_\theta
\end{array}
\right) 
&=& \frac{\muo I \, \sin\alpha}{ 2 a} \sum_{n=1}^\infty   \! \left(\frac{r}{a}\right)^{\!n-1} P_n^1(\cos\alpha) \nonumber \\
&& \times 
\left( 
\begin{array}{c}
 -P_n(\cos\theta)  \\[.2cm]
P_n^1(\cos\theta)
\end{array}
\right) \, .
\label{Bloop}
\end{eqnarray}
All that remains is to multiply each $n$ component of the field by the
appropriate shielding factor from Sec.~\ref{sec:multiplespheres} to
determine the interior field.  Furthermore, the reflected exterior
field, dominated by the response of the outermost magnetic shield, may
be determined by applying the results of Sec.~\ref{sec:singlesphere} to
each $n$ component.  An appropriate sensor and coil system to
effectively cancel particularly problematic external dipoles of this
sort can then be devised.

\subsection{Generation of a uniform internal field}
\label{internalCoils}

A critical requirement of many
experiments~\cite{bib:gpe1,bib:gpe2,bib:gpe3,bib:gpe4} is the
generation of a highly uniform magnetic field in the inner volume of a
passive shield system.  If the coils used to generate this field are
not \textit{self-shielded} in some manner~\cite{cb} they will couple
strongly to the shields.  This coupling, if taken into account
properly, can be used advantageously to improve the field homogeneity
over the case where no passive shielding is present.  In order to
accentuate this point, and to illustrate the power of our formulation,
we present two simple, canonical examples -- the saddle-shaped coil
and the Helmholtz coil.

\begin{figure}[hbt]
\centering
\includegraphics[keepaspectratio, width=3.3in]{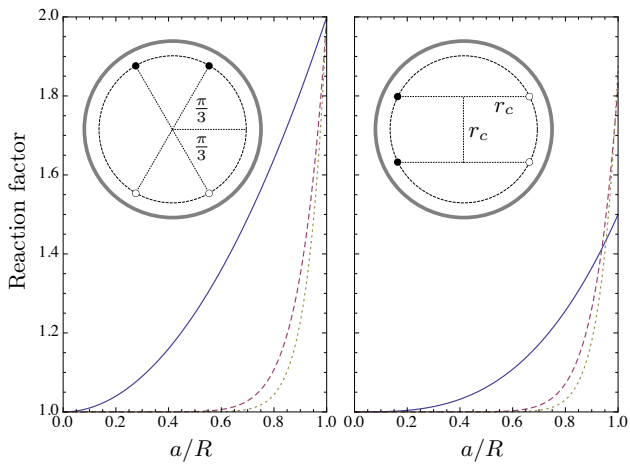}
\caption{The reaction factor for $n=1$ (solid line), 5 (dashed
  line), and 7 (dotted line) from Eqs.~\ref{Ccyl} and~\ref{Csphere}
  for the cylindrical (left) and spherical (right) case, respectively.
  Insets: Schematic of a saddle coil (left) and Helmholtz coil (right)
  located on the radius $a$ (dashed line) inside a shield (thick gray
  line) of inner radius $R$.  The dotted lines define the geometry of
  the coil and give the locations of the current (circles). The closed
  (opened) symbols indicate current flow out of (into) the page. }
\label{Cplots}
\end{figure}

In the cylindrical case, a saddle-shaped coil can be used to produce a
transverse field with a dominant $n=1$ term near $\rho =0$.  For a
very long (infinite) coil, the $n=2$ term can be eliminated by placing
the four axial current paths at $\phi = \pm \tfrac{\pi}{3}$ and $ \pm
\tfrac{2\pi}{3}$.\cite{cb} An end view of the geometric arrangement of
the currents is shown in Fig.~\ref{Cplots}.  Using the results of
Ref.~\cite{cb} along with Eq.~\ref{Ccyl} from above, the components of
the internal field of such a coil inside a high-$\mu$ cylindrical
shield are
\begin{eqnarray}
\left( 
\begin{array}{c}
B_\rho \\[.2cm]
B_\phi
\end{array}
\right) 
&=& \frac{2\muo I}{ \pi a} \sum_{n=1,5,7,\dots}^\infty   \!\sin \! \left(n\tfrac{\pi}{3}\right) \left(\frac{\rho}{a}\right)^{n-1}  \nonumber \\
&& \times \left[ 1+ \left( \frac{a}{R} \right)^{2n}\right] 
\left( 
\begin{array}{c}
\cos n\phi   \\[.2cm]
- \sin n\phi
\end{array}
\right) \, ,
\end{eqnarray}
where the sum is over odd $n$ not equal to an integer multiple of 3.
Provided that the coil is not located directly on the inner surface of
the shield (i.e., $a<R$) the resulting field is always more
homogeneous than an unshielded coil ($R\rightarrow \infty$), because
the term in square braces -- the reaction factor -- is greatest for
$n=1$ and decreases for all higher order terms.~\cite{superconductor}

As can be seen from the plot in Fig.~\ref{Cplots}, however, there must
exist a value of $a/R$ for which the ratio of the reaction factor for
$n=5$ compared to that for $n=1$ is a minimum.  One can show that this
occurs at $a/R=0.7784$ and that for a coil located at this position
the $n=5$ term is $\sim 33$\% lower relative to the $n=1$ term then
compared to the unshielded case.  This result demonstrates that field
homogeneity can be obtained not only by appropriate coil design but
also by a judicious choice of $a$ for the location of the coil inside
the shield.

Turning to the spherical case, a Helmholtz coil an be used to produce
an axial field with a dominant $n=1$.  The coil is constructed from
two current loops located at $z=\pm r_c/2$ (as shown in
Fig.~\ref{Cplots}), or equivalently at angles $\alpha$ and
$\pi-\alpha$ where $\sin\alpha=2/\sqrt{5}$ and
$\cos\alpha=1/\sqrt{5}$.  Since $\sin(\pi-\alpha)=\sin\alpha$,
$\cos(\pi-\alpha)=\cos\alpha$, and $P_n^1(u)$ is an even (odd)
function of $u$ for odd (even) degree $n$, only the odd $n$ terms of
Eq.~\ref{Bloop} contribute to net field.  Furthermore, since
$P_3^1(\pm \small{1/\sqrt{5}})$ is uniquely zero -- thereby
eliminating the $n=3$ term -- the field components can be written as
\begin{eqnarray}
\left(
\begin{array}{c}
B_r \\[.2cm]
B_\theta
\end{array}
\right) 
&=& \frac{2\muo I }{ \sqrt{5} \, a} \, \sum_{n=1,5,7,\dots}^\infty   \! \left(\frac{r}{a}\right)^{\!n-1} P_n^1(1/\sqrt{5}) \nonumber \\
&& \times 
\left( 
\begin{array}{c}
 -P_n(\cos\theta)  \\[.2cm]
P_n^1(\cos\theta)
\end{array}
\right) \, .
\label{BHelm}
\end{eqnarray}
The expansion of Eq.~\ref{BHelm} in $r=z$ at $\theta=0$ gives the
following leading order terms -- corresponding here to $n=1$ and~5 --
for the field along the central axis:
\begin{equation}
B_z= B_c \! \left(1-\frac{144}{125} \left(\frac{z}{r_c}\right)^4 +
\dots \right) \, ,
\label{BHelm0}
\end{equation}
where $B_c=\muo I/ r_c \times ( 4/5)^{3/2}$ is the well-known central field of a Helmholtz coil.

If the coil is now placed inside a high-$\mu$ spherical shield of
inner radius $R>a$, the relative strength of these terms will vary
according to Eq.~\ref{Csphere}, and as plotted in Fig.~\ref{Cplots}.
One can show that for $a/R=0.7817$ the ratio of the reaction factor
for $n=5$ compared to that for $n=1$ is a minimum and the relative
strength of the $z^4$ term is reduced by $\sim 15$\% compared to the
unshielded coil.  For $a/R>0.9381$, where the reaction factor of the
$n=5$ term becomes greater than that for $n=1$, the homogeneity near
the origin is in fact degraded.  This highlights the care that must be
taken in designing shield-coupled coils, even in ideal geometrical
situations.

\section{Conclusion}

In this paper, we have provided a comprehensive framework of
analytical results that are useful to analyze magnetic shielding
systems possessing cylindrical or spherical symmetry.  The results are
general to any shields possessing this symmetry, placed in any
exterior or interior dc current distribution that can be decomposed
into multipoles.  We have provided here but a few examples
demonstrating the utility of the approach, using our formulae to
analyze coil systems placed close to magnetic shields.

A key general finding is that higher order multipoles $n$ are always
shielded progressively better than the uniform field $n=1$ case.
Furthermore, judicious choices in geometry can make good designs of
homogeneous internal coil systems even better when placed internal to
a system of magnetic shields.  But the work goes beyond these findings
in allowing general magnetic shielding and coil system designs to be
studied.

In future work, we intend to study the deviations from these results
when more realistic geometries are considered, such as finite
cylindrical shields with end caps and apertures for experimental
access.  Such problems do not generally afford analytic solutions, and
one naturally must resort to finite element analysis (FEA) codes to
conduct such a study.  As a result, we envision first benchmarking FEA
code to the analytic formulae provided here for an ideal case
approximating the eventual desired coil and shield system.  Changes
are then made in the FEA model to represent more realistic coils
and/or shields, and the deviation from the ideal case quantified.  We
believe that in such a manner our results presented here will be
applicable to broad classes of coil and shielding systems, as an
important starting point toward achieving design goals.



%
%

%

\begin{acknowledgments}
We gratefully acknowledge the support of the Natural Sciences and Engineering Research Council of Canada. 
\end{acknowledgments}




\begin{thebibliography}{00}
\bibitem{code}An example of such code, written in Mathematica, is available from the authors upon request.
\bibitem{superconductor}  It is informative to contrast this with the case
of a superconducting or perfectly conducting shield, where free
currents induced on the inner surface of the shield act to reduce the
field in the region $\rho<a$.  In this case, the term in square braces is replaced by $[1- (a/R)^{2n}]$ (as given in Eq.~31 of Ref.~\cite{cb}) and the field becomes less homogeneous in the presence of the shield.
\end{thebibliography}
\end{document}